\documentclass[twocolumn,showpacs,amsfonts,amsmath,amssymb,superscriptaddress]{revtex4}
\usepackage{graphicx}
\usepackage{dcolumn}
\usepackage{bm}

\begin{document}

\title{Experimentally obtaining the Likeness of Two unknown qubits\\ on an NMR Quantum Information Processor}

\affiliation{Department of Modern Physics, University of Science
and Technology of China,
  Hefei, People's Republic of China, 230027\\}
\affiliation{Laboratory of Structure Biology, University of
Science and Technology of China,
  Hefei, People's Republic of China, 230027}

\author{\surname{XUE} Fei}
\email{Feixue@mail.ustc.edu.cn}
\author{\surname{DU} JiangFeng}
\author{\surname{ZHOU} XianYi}
\author{\surname{HAN} RongDian}
\affiliation{Department of Modern Physics, University of Science
and Technology of China,
  Hefei, People's Republic of China, 230027\\}
\author{\surname{WU} JiHui}
\affiliation{Laboratory of Structure Biology, University of
Science and Technology of China,
  Hefei, People's Republic of China, 230027}

\date{\today}

\begin{abstract}
Recently quantum states discrimination has been frequently
studied. In this paper we study them from the other way round, the
likeness of two quantum states. The fidelity is used to describe
the likeness of two quantum states. Then we presented a scheme to
obtain the fidelity of two unknown qubits directly from the
integral area of the spectra of the assistant qubit(spin) on an
NMR Quantum Information Processor. Finally we demonstrated the
scheme on a three-qubit quantum information processor. The
experimental data are consistent with the theoretical expectation
with an average error of $0.05$, which confirms the scheme.
\end{abstract}


\maketitle

Quantum Information Processing(QIP) has been the subject of much
recent interest, not only because it has great advantages in
efficient algorithms and secure communications, but also because
quantum information differs from classical information in several
fundamental ways. One important difference is that qubits can
hold superposition states while classical bits can only hold
either 0 or 1 at the same time. So unlike two classical bits
whose relationship is either the same or inverse, the
relationship of two qubits is more complex. The complication
comes from the superposition principle of quantum mechanics.
Recently quantum state discrimination has been studied
frequently.\cite{Lee,Shengyu,Barnett,Chefles,Walgate} In this
paper we study quantum states from the other way round, i.e.,
considering the likeness of two quantum states. We focus on such
questions that whether two unknown quantum states are the same or
not, morever, the extent to which two states are alike, which is
useful in quantum encryption and quantum states comparison.

Since Gershenfeld and Chuang realized quantum computation with NMR
technique in 1997\cite{Gershenfeld} a lot of jobs have been done
with Nuclear Magnetic Resonance Quantum Information Processor(NMR
QIP). NMR QIP has successfully demonstrated some efficient
algorithms, such as Deutsch-Jozsa algorithm\cite{DJ}, searching
algorithm\cite{Searching1,Searching2}, Bernstein-Vazirani parity
problem\cite{BV},  and Shor's quantum factoring
algorithm\cite{LMKVshor}; and some fundamental ideas in quantum
information, such as Dense Coding\cite{Dense}, error
correction\cite{Error}, quantum games\cite{Qgame}, creation of
Greenberger Horne Zeilinger states\cite{GHZ} and approximate
quantum cloning\cite{Clone}. In this paper we will enlarge the
list.

In the paper, first we give a brief review of the concept of the
fidelity which originated from quantitative measures of the
accuracy of transmission in communication theory, then we present
a scheme to obtain the quantitative likeness(the fidelity) of two
unknown qubits on an NMR QIP. The fidelity comes from the integral
area of the spectra of the assistant qubit(spin), rather than from
the tomography of the spin system. This makes it convenient to
implement the scheme by NMR QIP. Finally we demonstrated the
scheme on a three-qubit quantum information processor by obtaining
the fidelity of two unknown qubits.

Before explainning the scheme, let us first give a review of the
problem to be considered. The question is: given two unknown
quantum states $\vert  \psi_1 \rangle$ and $\vert  \psi_2
\rangle$, what is the relationship of them or how similar they
are. In order to quantitively describe the likeness some functions
are needed.

Naturally we want that the function has the following characters:

a. $0\leq F(\psi_1,\psi_2)\leq 1$ and $F(\psi_1,\psi_2)=1$, if and
only if $\vert  \psi_1 \rangle=\vert  \psi_2 \rangle$.

b. $F(\psi_1,\psi_2)=F(\psi_2,\psi_1)$.

c. $F(\psi_1,\psi_2)$ is invariant under any unitary
transformations on both states.

The fidelity appearing in the communication theory is a good
candidate for it. The origin of the fidelity is a quantitative
measure of the accuracy of transmission. It has desired properties
and thus is a sensible choice as the quantitative measure of the
likeness of two unknown quantum states. The fidelity of two
quantum states is defined as \cite{Jozsa}
\begin{equation}\label{fidelityofpure}
F(\psi_1,\psi_2)=\vert\langle\psi_1\vert\psi_2\rangle\vert^2.
\end{equation}
Now we have the function of the quantitative measure of the
likeness of two quantum states, then the question turns to be
obtaining the fidelity $F(\psi_1,\psi_2)$.

\begin{figure}[tph]
\includegraphics{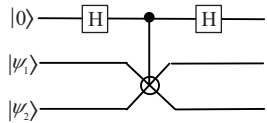}
\caption{\label{network} Quantum circuit to obtain the likeness of
two unknown quantum states $\vert  \psi_1 \rangle$ and $\vert
\psi_2 \rangle$.}
\end{figure}

In Quantum Fingerprinting\cite{fingerprinting}, Harry Buhrman have
presented a quantum circuit (Fig.\ref{network}). It was used to
test $\vert  \psi_1 \rangle= \vert  \psi_2 \rangle$ or
$\vert\langle\psi_1\vert\psi_2\rangle\vert\leq\delta$ in that
paper. We found that this quantum circuit can be further used to
obtain fidelity $F(\psi_1,\psi_2)$ of the two unknown quantum
states.
\begin{equation}\label{eqnetwork}
(H\otimes I)(G_{Fredkin})(H\otimes
I)\vert0\rangle\vert\psi_1\rangle \vert\psi_2\rangle
\end{equation}
where H is the Hadamard gate, which maps $\vert
b\rangle\rightarrow\frac{1}{\sqrt2}(\vert0\rangle +
(-1)^b\vert1\rangle)$. $G_{Fredkin}$ is the Fredkin gate which is
also called contolled-swap gate, the first qubit is the
controlling qubit, the controlled qubits can be generalized to
quantum states. Tracing through the execution of the circuit, the
final state is
\begin{equation}\label{finals}
\frac{1}{2} \vert 0 \rangle ( \vert \psi_1 \rangle \vert \psi_2
\rangle + \vert \psi_2 \rangle \vert \psi_1 \rangle ) +
\frac{1}{2} \vert 1 \rangle ( \vert \psi_1 \rangle \vert \psi_2
\rangle - \vert \psi_2 \rangle \vert \psi_1 \rangle )
\end{equation}
After tracing out the other two qubits, the density matrix of the
first qubit of this final state is reduced to the form,
\begin{equation}\label{finalo1}
\begin{pmatrix}
  \frac{1}{2}(1+\vert\langle\psi_1\vert\psi_2\rangle\vert^2) & 0 \\
  0 &\frac{1}{2}(1-\vert\langle\psi_1\vert\psi_2\rangle\vert^2)
\end{pmatrix}
\end{equation}
When this state is projected to $\vert0\rangle$ and
$\vert1\rangle$, the difference of the two probability is
$\vert\langle\psi_1\vert\psi_2\rangle\vert^2$, which is just the
fidelity $F(\psi_1,\psi_2)$. Then the question is to obtain the
difference of two outcome probability.

\begin{figure}[tp]
\centering
\includegraphics{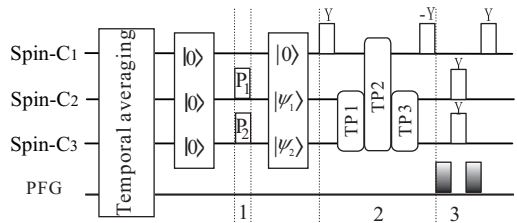}
\caption{\label{allpulse} The pulse sequence for the network of
obtaining the likeness of two qubits and the reading pulse
sequence. The symbols on the indicate the phases in which the
pulses are applied. All experiments were done on Bruker Avance
DMX400 spectrometer at temperature 300K.}
\end{figure}

The outcome probabilities can not be obtained from few experiments
for a general quantum computer. But for NMR QIP, which is a Bulk
Quantum computer\cite{Gershenfeld}, the situation is very
different. In NMR QIP pseudo-pure states which are deviation
density matrices of the spin system were proposed as the initial
states, in stead of genuine pure states\cite{Gershenfeld,Ppure}.
The system of the spins in NMR may be convenient described by the
product operator notation\cite{OW}. The state of the first qubit
can sure be expressed in the notation of the deviation density
matrix and the notation of product operator,
\begin{eqnarray}
\rho_1=&&
\begin{pmatrix}
  \rho_{00} & \rho_{01} \\
  \rho_{10} & \rho_{11}
\end{pmatrix} \\ \nonumber
= &&(\rho_{00}- \rho_{11}) I_z^1
+ (\rho_{01}+\rho_{10})I_x^1 \\
&& +(\rho_{01}-\rho_{10})iI_y^1 + \frac{1}{2}(\rho_{00} +
\rho_{11})I
\end{eqnarray}

where $I_a^i=\frac{1}{2}\sigma_a^i$, $a=x, y ,z$, and  $I$ is the
identity matrix. So if the coefficient of $I_z^1$ could be
observed, then the difference of the two probabilities--the
fidelity is obtained. Of course it can be done by constructing the
density matrix of the spin system. However the fidelity can be
gotten in a more simple method. The fidelity--the coefficient of
$I_z^1$ can be gotten by the following
operations(Fig.\ref{allpulse} part 3). First apply a gradient
pulse to remove the non-diagonal part of the density matrix if
there are any, then get rid of the part that correspond to the
spins beside the first spin, finally apply a $90_y$ pulse on the
first qubit. After performing these operations the state of the
spin system becomes
\begin{eqnarray}
 &&(\rho_{00}- \rho_{11}) I_x^1 + \frac{1}{2}(\rho_{00} +
 \rho_{11})I.
\end{eqnarray}
The identity matrix in NMR is not observable, so the signal from
the first spin, the integral area of peaks, now corresponds to
$\rho_{00}- \rho_{11}$, which is proportional to
$\vert\langle\psi_1\vert\psi_2\rangle\vert^2$-the fidelity
$F(\psi_1,\psi_2)$.

On a three-qubit alanine(in $D_2O$) NMR QIP {\cite{FredkingG}}, we
demonstrated the scheme. Spin-$C_1$ serves as qubit-1(the
assistant spin), Spin-$C_2$ serves as qubit-2, which holds $\vert
\psi_1 \rangle$, and Spin-$C_3$ serves as qubit-3, which holds
$\vert  \psi_2 \rangle$.  We let $\vert \psi_1
\rangle=\cos\frac{\theta_1}{2} \vert 0 \rangle - e^{i\varphi_1}
\sin\frac{\theta_1}{2} \vert 1 \rangle$ and $\vert \psi_2
\rangle=\cos\frac{\theta_2}{2} \vert 0 \rangle - e^{i\varphi_2}
\sin\frac{\theta_2}{2} \vert 1 \rangle$. Specifically we studied
two situations systemically: $\vert \psi_2 \rangle=\vert 0
\rangle$, changing $\vert  \psi_1 \rangle$ and $\vert  \psi_1
\rangle=\vert 0 \rangle$, changing $\vert  \psi_2 \rangle$. In
each situation we have collected 20 experimental data, where
$\theta_i$ varies from 0 to 180 degree, with a step of 45 degree,
and for each $\theta_i$, $\varphi_i$ varies as 0, 90, 180 and 270
degree.

The pseudo-pure state is implemented by temporal averaging of
three separate experiments, see Ref.\cite{fei} for the detail. The
pulse sequence of the scheme is shown in Fig.\ref{allpulse}. It is
applied after the preparation pulses for the pseudo-pure state. It
includes three parts: 1. Prepare $\vert \psi_1
\rangle=\cos\frac{\theta_1}{2} \vert 0 \rangle - e^{i\varphi_1}
\sin\frac{\theta_1}{2} \vert 1 \rangle$ or $\vert \psi_2
\rangle=\cos\frac{\theta_2}{2} \vert 0 \rangle - e^{i\varphi_2}
\sin\frac{\theta_2}{2} \vert 1 \rangle$, it is implemented by
selective pulses in certain phase on Spin-$C_2$ or Spin-$C_3$ . 2.
The sequence for network in Fig.\ref{network}. The two Hadamard
gates are implemented by $R_y^1(90)$ and $R_{-y}^1(90)$, the
Fredkin gate is implemented with three transition pulses
TP1,TP2,TP3, see Ref.\cite{FredkingG,Du2} for the detail. 3.
Reading sequence. Processing of tracing out Spin-$C_2$ and
Spin-$C_3$ is done by integrating the entire multiplet of the
Spin-$C_1$. The errors in the experiment are estimated by
$Err=\vert F_{exp} - F_{theory} \vert$.

\begin{figure}[tp]
\centering
\includegraphics{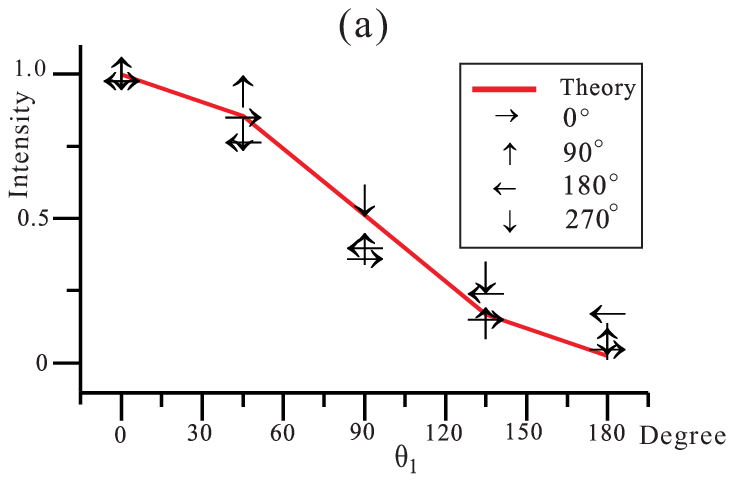}
\includegraphics{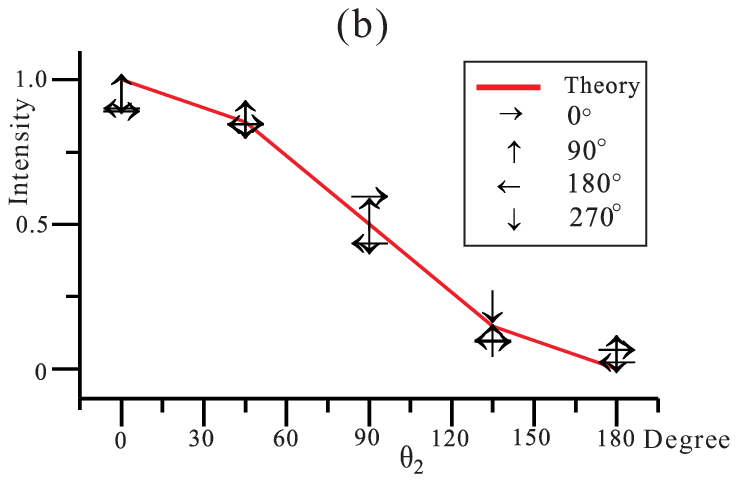}
\caption{\label{Eresult}Experiment results. Arrows are normalized
experimental data, directions of them indicate $\varphi_i$. $(a)$
corresponds $ \vert \psi_2 \rangle = \vert 0 \rangle$; $(b)$
corresponds $ \vert \psi_1 \rangle = \vert 0 \rangle$. The
theoretical fidelity
$F_{theory}(\psi_1,\psi_2)=cos^2(\frac{\theta_i}{2})$.}
\end{figure}

The experimental data each of which is the sum of three
experimental data, after the normalization, are plotted in
Fig.\ref{Eresult}. Though the biggest error in the experimental
data reaches about $0.11$. The experimental data are consistent
with the theoretical expectation with a average error of $0.05$.
It is sufficient to demonstrate the scheme. We have also tried the
experiments that both $\vert  \psi_1 \rangle$ and $\vert \psi_2
\rangle$ were changing. It is hard to make sure that both
$\varphi_1$ and $\varphi_2$ are what we specify, because that the
spins are precessing in different frequency, and the temporal
averaging technique is also increase the difficulty. If these are
taken into considering and be emendated the scheme can still be
confirmed. There are two main sources of the errors. One is the
imperfections of the pulses in the experiments, which bring about
some errors in the implementation of the circuit and can be
reduced by refining the pulse sequence. The other is the effect of
the relaxation times. The length of the pulse sequences reaches
$0.3s$, while $T_2$ of the alanine is about $0.893s$ in the
experiment. So the effect of the relaxation is not neglectable and
will influence the final spectra. Morever, different terms of the
spin system, such as $I_z$ and $I_x$, have different relaxation
behaviors and different relaxation speeds, hence these errors are
harder to be reduced.

To summarize, we introduced the fidelity as a quantitative measure
of the likeness of two unknown quantum states. Then we presented a
scheme to obtain the fidelity on an NMR Quantum Information
Processor, showed that the fidelity can be set proportional to the
intensity of the signal from the assistant qubit(spin). Finally we
experimentally demonstrated the scheme on a three-qubit NMR QIP
which is implemented by the solution of alanine. The
network(Fig.\ref{network}) with slight difference was studied in
Ekert's work\cite{Ekert,Entangle}. In a sense our work in this
paper gives an experimentally demonstration of some ideas in the
paper\cite{Ekert}. Besides on NMR QIP, any quantum computer based
on bulk spins can use the scheme to obtain the fidelity of two
unknown qubits.

This work was supported by the National Nature Science Foundation
of China (Grants No. 10075041, No. 10075044 and No. 10104014), and
the National Fundamental Research Program(Grant No. 2001CB309300).

\end{document}